\documentclass[aps,prb,amsmath,amssymb,twocolumn,floatfix,10pt,superscriptaddress,showpacs]{revtex4-1}

\usepackage{graphicx}
\usepackage{hyperref}
\usepackage{color}

\newcommand{\bs}{\;\;\;\;\;}

\newcommand{\ve}{\mathbf}

\newcommand{\D}{{\rm d}}

\newcommand{\E}{\mathrm{e}}
\newcommand{\I}{\mathrm{i}}

\begin{document}

\title{Effective models for strong electronic correlations at graphene edges}
\author{Manuel J. Schmidt}
\affiliation{Institut f\"ur Theoretische Festk\"orperphysik, JARA-FIT and JARA-HPC, RWTH Aachen University, 52056 Aachen, Germany}
\author{Michael Golor}
\affiliation{Institut f\"ur Theoretische Festk\"orperphysik, JARA-FIT and JARA-HPC, RWTH Aachen University, 52056 Aachen, Germany}
\author{Thomas C. Lang}
\affiliation{Department of Physics, Boston University, Boston, MA 02215, USA}
\author{Stefan Wessel}
\affiliation{Institut f\"ur Theoretische Festk\"orperphysik, JARA-FIT and JARA-HPC, RWTH Aachen University, 52056 Aachen, Germany}
\date{\today}


\begin{abstract}
We describe a method for deriving effective low-energy theories of electronic interactions at graphene edges. Our method is applicable to general edges of honeycomb lattices (zigzag, chiral, and even disordered) as long as localized low-energy states (edge states) are present. The central characteristic of the effective theories is a dramatically reduced number of degrees of freedom. As a consequence, the solution of the effective theory by exact diagonalization is feasible for reasonably large ribbon sizes. The quality of the involved approximations is critically assessed by comparing the correlation functions obtained  from the effective theory with numerically exact quantum Monte-Carlo calculations. We discuss effective theories of two levels: a relatively complicated fermionic edge state theory and a further reduced Heisenberg spin model. The latter theory paves the way to an efficient description of the magnetic features in long and structurally disordered graphene edges beyond the mean-field approximation.
\end{abstract}

\maketitle

\section{Introduction}

The Coulomb repulsion among electrons can give rise to exotic phenomena in solid-state materials if this interaction is sufficiently strong, subsumed under the term {\it strong correlations}. In graphene, the  honeycomb lattice of carbon atoms,\cite{novoselov_2004,peres_rmp_graphene_2010} the question of the strength of the Coulomb interaction is, however, not settled yet. On one hand there is a strong structural confinement to two dimensions, usually enhancing electronic correlations. But on the other hand, the relativistic dispersion near the Dirac points with its vanishing density of states (DOS) at the charge neutrality point tends to suppress correlation effects. The bulk of graphene seems to be on the verge  of the critical interaction strength beyond which interaction-driven phase transitions set in.\cite{meng_spin_liquid_2010,Sorella12,kotov_rmp_interactions_2012,wehling_red_U_2013,drut1,drut2}

At the edges of a honeycomb-lattice nanostructure, the situation may be significantly different. While the strong electronic confinement to a two-dimensional plane persists, there is no constraint on the DOS to remain small at an edge. Instead, it happens that peaks in the DOS emerge, depending on details of the edge geometry. Within such spectral ranges, the Coulomb repulsion becomes important and leads to interaction-induced phenomena, such as edge magnetism at zigzag edges.\cite{wakabayashi_edge_magnetism_1998} Edge magnetism has been extensively studied theoretically for clean zigzag edges with a great variety of methods ranging from self-consistent field theories to quantum Monte Carlo (QMC) simulations (see, e.g., Refs. \onlinecite{zigzag1,zigzag2,zigzag3,zigzag4,zigzag5,zigzag6,zigzag7,zigzag8,zigzag9,schmidt_tem_2010,luitz_ed_2011}). But edge magnetism is not restricted to zigzag edges. Also in chiral ribbons edge magnetism was examined theoretically.\cite{yazyev_chiral_2011,golor_2013} Recently experimental evidence for spin-polarized edge states has been reported in Ref.~\onlinecite{crommie_2011}. But it also turned out that in realistic situations the magnetic properties of graphene edges depend on the environment (passivation, substrate).\cite{yan_gnr_on_substrates_2012} In particular it was found that the edge states hybridize strongly with the surface state of an Ir(111) surface, which effectively leads to their destruction.

Most theoretical studies of edge magnetism have been performed numerically in the basis of the carbon $p_z$-orbitals arranged in a honeycomb lattice. Such calculations are typically restricted to relatively small systems with a few ten-thousand lattice sites for, e.g., mean-field calculations, or even to below thousand sites for quantum Monte-Carlo or {\it ab initio} methods.

Here, we follow a different approach. Before we perform the actual calculations, we dramatically reduce the degrees of freedom to only the relevant ones, namely the edge states. Those have (nearly) zero energy and a strong confinement to the edge (in addition to the confinement to the 2D graphene plane). We derive an effective theory for edge states at general graphene edges (zigzag, chiral, disordered). For the remaining states (called bulk states) it turns out that they can be dropped from the effective theory. For observables such as the spin-spin correlation function on the lattice, some corrections from the bulk states must be taken into account finally. The crucial point is that all non-perturbative effects (such as long-range correlations) are contained in the effective theory for the edge states. The residual bulk states may be accounted for within perturbation theory or even in a non-interacting approximation. Their only effect is a trivial local amplification of the magnetic correlations induced by the edge states.

This approximation, i.e., the reduction to an effective theory for the edge states only, has been used before,\cite{zigzag5,schmidt_tem_2010,luitz_ed_2011} but in this paper we critically assess the quality of this approximation by comparing the results of the effective low-energy theory with numerically exact quantum Monte-Carlo simulations. We find that the quantitative accuracy which can be reached within this effective theory is at least as good as the limits set by uncertainties in the parameters entering the calculations (i.e., the hopping amplitude $t$ and the Hubbard parameter $U$). The relevant qualitative behavior, i.e., the long-range structure of the spin-spin correlations, is remarkably well reproduced.

Even more important than the high quality of the approximation is the fact that the effective theory separates the relevant degrees of freedom from the irrelevant ones, and thereby providing valuable insights into the underlying physics. The corrections from the bulk states, mentioned above, actually tend to screen the non-trivial edge state effects so that it is sometimes difficult to extract the underlying physics, e.g., from the exact results of a quantum Monte-Carlo simulation of the full lattice model. For instance, the existence of edge magnetism is difficult to extract from the exact correlation functions for certain chiral ribbons,\cite{golor_2013} since it is not always clear if a non-zero correlation function is a bulk effect that decays as a power law in the thermodynamic limit, or if it is an edge state effect that gives rise to long-ranged correlations.
The effective theory discussed here does not suffer from these issues, as it provides direct access to the part of the system that potentially 
leads to long range correlations and separates this part from the obfuscating bulk of the system.

This paper is organized as follows. In Section \ref{sec_fermionic_theory}, we describe the effective edge state theory and the different contributions to the spin-spin correlation function on the honeycomb lattice. In Section \ref{sec_qmc}, the spin-spin correlation function calculated within the effective theory is benchmarked against numerically exact quantum Monte-Carlo (QMC) results. An intuitive picture of the basic mechanisms important for the magnetic correlations is developed in Sec. \ref{sec_intuitive_picture}. Finally, in Sec. \ref{sec_heisenberg}, we perform a further approximation, arriving 
at an effective Heisenberg model for chiral nanoribbons, and show that the magnetic correlations in the more complicated fermionic theory for the edge states are  reproduced.

\section{Effective fermionic theory for localized edge states\label{sec_fermionic_theory}}

\subsection{Derivation of the effective theory\label{sec_derivation}}

We start from the nearest neighbor hopping Hamiltonian of a general honeycomb lattice
\begin{equation}
H_0 = \sum_{\langle i,j\rangle,\tau} c^\dagger_{i\tau} c_{j\tau} + \mathrm{H.c.},
\end{equation}
where $c_{i\tau}$ annihilates an electron with spin $\tau$ at site $i$ of the lattice and $\langle i,j\rangle$ runs over nearest neighbors. Let $N$ be the total number of lattice sites. $H_0$ can be rewritten in terms of its exact eigenstates $d^\dagger_{\mu \tau} = \sum_i \psi_\mu(i) c^\dagger_{i\tau}$, with $\psi_\mu(i)$ the normalized single-particle wave function and $\mu$ a (collective) index,
\begin{equation}
H_0 = \sum_{\mu,\tau} \epsilon_\mu d^\dagger_{\mu\tau} d^{}_{\mu\tau},\label{orig_hopping}
\end{equation}
where $\epsilon_\mu$ denotes the associated eigenenergies.

The Hubbard Hamiltonian $H_U = U \sum_{i}c^\dagger_{i\uparrow}c^{}_{i\uparrow}c^\dagger_{i\downarrow}c^{}_{i\downarrow}$ may be written in the eigenbasis of $H_0$
\begin{equation}
H_U = U \sum_{\mu_1\dots\mu_4} \Gamma_{1234} d^\dagger_{\mu_1\uparrow} d^{}_{\mu_2\uparrow}d^\dagger_{\mu_3\downarrow} d^{}_{\mu_4\downarrow},\label{orig_hubbard}
\end{equation}
with
\begin{equation}
\Gamma_{1234} = \sum_i \psi_{\mu_1}^*(i)\psi^{}_{\mu_2}(i)\psi_{\mu_3}^*(i)\psi^{}_{\mu_4}(i).\label{gamma_def}
\end{equation}

The central step in the derivation of the effective theory is the partitioning of the eigenstates of $H_0$ into edge states and bulk states
\begin{equation}
\{d_{\mu\tau}\} = \{e_{\rho\tau}\}\cup \{b_{\nu\tau}\} .
\end{equation}
For a single zigzag edge this partition is obvious: states with zero energy are edge states and those with finite energy are bulk states. For arbitrarily-shaped graphene structures, however, this energetic criterion is too simplistic for a reasonable separation of bulk and edge states. Instead, one may distinguish edge from bulk states by the maximum of the wave function weight $l_\mu=\max_i |\psi_\mu(i)|^2$, which is a convenient localization measure. If $l_\mu$ is of order $N^{-1}$, the state $\mu$ is a bulk state, otherwise it is an edge state. For a conventional zigzag edge, for instance, $l_\mu$ is of the order of the inverse length of the edge. For finite-size systems the partition into edge- and bulk states is not rigorous (we will see examples later where it is not clear if a certain eigenstate of $H_0$ should be labeled as a bulk state or an edge state). In practice, however, it is always possible to obtain a feasible partition. If in doubt, one may always include more states in the edge-state set. 
In the worst case, if a bulk state is included erroneously in the edge-state set, the Hilbert space of the effective theory becomes somewhat larger than necessary -- the physics however remains unchanged.

With this separation into edge and bulk states, we may express the Hubbard Hamiltonian as
\begin{multline}
H_U = U \sum_{1234} \Gamma_{1234} e^\dagger_{1\uparrow} e^{}_{2\uparrow}e^\dagger_{3\downarrow} e^{}_{4\downarrow} \\+ U \sum_{1234} \Gamma_{1234} \sum_\tau e^\dagger_{1\tau} e^{}_{2\tau} b^\dagger_{3\bar\tau} b^{}_{4\bar\tau} + \dots,\label{eff_hubbard}
\end{multline}
where the dots refer to terms containing an odd number of bulk state operators or four bulk state operators. Here and henceforth, an overbar of a binary index, such as edge or spin, denotes inversion, i.e. $\tau$ and $\bar \tau$ are opposite spins. In order to shorten the notation, we use the same numeric index symbols for edge and bulk states. Whether an index, e.g., in a vertex function $\Gamma_{1234}$ corresponds to a bulk state or to an edge state can always be determined from the corresponding operators. For instance, in $\Gamma_{1234}e^\dagger_{1\tau}e^{}_{2\tau}e^\dagger_{3\tau'}b^{}_{4\tau'}$ the indices 1,2,3 run over the edge states while index 4 runs over the bulk states. If a vertex function or an energy is written without their corresponding electron operators, the bulk/edge indices will be indicated by $b/e$ superscripts ($\Gamma_{1234}^{eeeb}$ in the example above).

Up to now we have only expressed the Hamiltonian in a different basis. The first approximation we make in approaching an effective low-energy theory is to neglect all but the first two terms in Eq.~(\ref{eff_hubbard}). We have also performed a more controlled approximation based on a Schrieffer-Wolff transformation to eliminate the leading order of the terms omitted in Eq.~(\ref{eff_hubbard}). The resulting additional fermionic couplings were extremely small, however, so that  their effect could not be observed in all geometries discussed in this work. Thus, we have chosen to drop those second-order terms completely and keep only the terms given in Eq. (\ref{eff_hubbard}). The second approximation consists in replacing the remaining two bulk state operators by their average with respect to the non-interacting Slater-determinant
\begin{equation}
b^\dagger_{1\tau}b^{}_{2\tau} \approx \delta_{12} \Theta(-\epsilon^b_{1}).
\end{equation}
Thus, the second term in Eq. (\ref{eff_hubbard}) involves an effective edge state hopping $t^*_{12}e^\dagger_{1\tau}e^{}_{2\tau}$ with
\begin{equation}
t^*_{12} = U \sum_{3} \Gamma^{eebb}_{1233} \Theta(-\epsilon^b_3) = U \sum_i [\psi^e_1(i)]^*\psi_2^e(i) \rho_b(i),
\end{equation}
where $\rho_b(i)$ is the electronic density at lattice site $i$ derived from all occupied bulk states. As was shown in Appendix B of Ref. \onlinecite{schmidt_tem_2010}, $\rho_b(i)$ can be calculated directly from the total edge state density $\rho_e(i) = \sum_1 |\psi_1^e(i)|^2$ by $\rho_b(i) = \frac12 (1-\rho_e(i))$ as long as particle-hole symmetry is present.\footnote{Note that even if particle hole symmetry is slightly broken, e.g., by second nearest neighbor hoppings, this simplification is still an excellent approximation.} The site-independent part of $\rho_b$ may be dropped as it only leads to a chemical potential term which is usually compensated by the positive background charge from the lattice. In the particle-hole symmetric case, one may furthermore write $\frac12 \rho_e(i) = \sum_{34} \psi_3^*(i) \psi^{}_4(i) \langle e^\dagger_{3\tau} e^{}_{4\tau} \rangle_0$, where $\langle\cdot\rangle_0$ is the average with respect to the non-interacting ground state in which exactly half of the edge states are filled. 
Therefore, the effective hopping may be conveniently absorbed into the Hubbard part of the effective Hamiltonian
\begin{equation}
H_{\rm eff} = \sum_{\mu\tau} \epsilon_\mu e^\dagger_{\mu\tau}e^{}_{\mu\tau} + U \sum_{1234} \Gamma_{1234} :e^\dagger_{1\uparrow}e^{}_{2\uparrow} : :e^\dagger_{3\downarrow} e^{}_{4\downarrow}:,
\end{equation}
where $:e^\dagger_{1\tau}e^{}_{2\tau}: = e^\dagger_{1\tau}e^{}_{2\tau} - \langle e^\dagger_{1\tau}e^{}_{2\tau}\rangle_0$.

The approximations made above are essentially based on the assumption that the electronic correlations within the bulk states and also the cross correlations of edge and bulk states are negligible. In fact, this statement will be slightly relaxed subsequently as it turns out that including the bulk background susceptibility enhances the agreement of the correlation functions calculated within QMC and the effective theory. Nevertheless, the nontrivial long-range physics is completely contained in $H_{\rm eff}$ as long as $U$ is below its bulk critical strength beyond which the whole graphene system (and not only its edge) is in an ordered phase. These {\it background corrections} only lead to an enhancement in the correlation function, the basic structure of the correlation function is still solely determined by the edge states.

\subsection{Correlation functions on the lattice}

By now we have derived an effective low-energy theory for edge states. For calculating correlation functions on the original lattice (this is needed, e.g., for benchmarking against QMC calculations), we need to translate the effective-theory correlation functions back to the original formulation on the honeycomb lattice. We consider the spin-spin correlation function 
\begin{equation}
\langle \sigma_i^z\sigma_j^z\rangle = \sum_{\tau\tau'}\tau\tau' \langle c^\dagger_{i\tau}c^{}_{i\tau} c^\dagger_{j\tau'}c^{}_{j\tau'}\rangle,
\end{equation}
formulated in terms of the lattice operators $c_{i\tau}$. These may be transformed to a new basis $d_{\mu\tau}$ of $H_0$ eigenstates so that
\begin{equation}
\langle \sigma_i^z\sigma_j^z\rangle = \sum_{\substack{1234\\\tau\tau'}} \tau\tau' \psi^*_1(i)\psi^{}_2(i) \psi^*_3(j)\psi^{}_4(j) \langle d_{1\tau}^\dagger d^{}_{2\tau}d^\dagger_{3\tau'}d^{}_{4\tau'}\rangle.\label{corr_d}
\end{equation}
As in the derivation of the effective theory, we partition the $d_{\mu\tau}$ basis states into bulk states $b_{\nu\tau}$ and edge states $e_{\rho\tau}$. Such a bipartition leads to a large number ($2^4$) of different terms in Eq. (\ref{corr_d}). However, only few of them give rise to significant contributions in the actual correlation function. We begin by neglecting all couplings between edge and bulk states, for the same reasons as discussed in the previous section. Thus, the only non-vanishing terms are
\begin{align}
\langle \sigma^z_i\sigma^z_j\rangle_{e} &= \sum_{\substack{1234\\\tau\tau'}} \tau\tau' \psi^*_1(i)\psi^{}_2(i) \psi^*_3(j)\psi^{}_4(j) \langle e_{1\tau}^\dagger e^{}_{2\tau}e^\dagger_{3\tau'}e^{}_{4\tau'}\rangle, \nonumber \\
\langle \sigma^z_i\sigma^z_j\rangle_{b} &= \sum_{\substack{1234\\\tau\tau'}} \tau\tau' \psi^*_1(i)\psi^{}_2(i) \psi^*_3(j)\psi^{}_4(j) \langle b_{1\tau}^\dagger b^{}_{2\tau}b^\dagger_{3\tau'}b^{}_{4\tau'}\rangle\nonumber \\
\langle \sigma^z_i\sigma^z_j\rangle_{m} &=  \sum_{\substack{1234\\\tau}} \psi^*_1(i) \psi^{}_2(j) \psi^{}_3(i) \psi^*_4(j) \langle b_{1\tau}^\dagger b^{}_{2\tau}\rangle \langle e^{}_{3\tau} e^\dagger_{4\tau}\rangle\nonumber \\
&\bs\bs\bs +[e\leftrightarrow b],
\end{align}
and originate from the correlations of the interacting edge states ($e$), the non-interacting bulk states ($b$), and a mixed term ($m$), involving non-local bulk state and edge state densities. In $\langle\sigma^z_i\sigma^z_j\rangle_{m}$ the symbol $[e\leftrightarrow b]$ stands for the first term with all $b$ and $e$ operators interchanged.

The bulk- and mixed terms will turn out to be important only for the sake of a quantitative comparison to the QMC correlation functions. They are relatively short-ranged compared to the edge state contribution and do not contain interesting physics. The edge state term, however, contains all the non-trivial effects related to long-range correlations and edge magnetism. These effects are contained in $\langle e^\dagger_{1\tau} e^{}_{2\tau} e^\dagger_{3\tau'} e^{}_{4\tau'}\rangle$, which will be evaluated by exact diagonalization in the present paper.

The three terms in the correlation function will be shown to be in good agreement with the QMC correlation functions at graphene edges for small $U$ and reproduce the transition from a local antiferromagnetic correlation to extended ferromagnetism along the ribbon edges. However, if $U$ becomes comparable to $t$, additional bulk corrections originating from the residual interaction matrix elements between edge and bulk states must be taken into account. This residual interaction on top of the edge state correlation function acts as an additional Zeeman field to which the bulk states respond linearly. This leads to a correction to the correlation function (for details see Appendix \ref{appendix_background_correction})
\begin{equation}
C^{(1)}(i,j) = U \sum_{i_1} \left[ \langle \sigma^z_i \sigma^z_{i_1}\rangle^e \chi_{i_1 j}^b + \chi_{i i_1}^b \langle \sigma^z_{i_1} \sigma^z_{j}\rangle^e\right],
\end{equation}
where
\begin{equation}
\chi^b_{ij} = \sum_{12}\frac{2 {\rm Re}[\psi^*_1(j)\psi^{}_1(i) \psi^{}_2(j) \psi^*_2(i) ]}{\epsilon_1-\epsilon_2} \Theta(\epsilon^b_1)\Theta(-\epsilon^b_{2})
\end{equation}
is the bulk state spin susceptibility. The sum only contains bulk state wave functions. The quality of the correction can be further increased by taking the interaction between the bulk states into account in the susceptibility. This may be done within random-phase approximation (RPA), i.e., by replacing
\begin{equation}
\chi^b \rightarrow \chi^{b,\mathrm{RPA}} = \frac{\chi^b}{1- U \chi^b},
\end{equation}
such that we denote the correlation function $C^\text{(RPA)}$, respectively. The corresponding background correction in RPA reads
\begin{equation}
C^{(\mathrm{RPA})}(i,j) = U \sum_{i_1} \left[ \langle \sigma^z_i \sigma^z_{i_1}\rangle^e \chi_{i_1 j}^{b,\mathrm{RPA}} + \chi_{i i_1}^{b,\mathrm{RPA}} \langle \sigma^z_{i_1} \sigma^z_{j}\rangle^e\right].
\end{equation}

\section{The quality of the effective model approximations\label{sec_qmc}}

\subsection{The quantum Monte Carlo method}
The effective model is tested against numerically exact results, obtained by projective auxiliary-field
determinant QMC calculations, in which the groundstate spin-spin correlations on finite ribbons are calculated
as
\begin{equation}
\label{eq:qmc}
\langle\sigma_i^z\sigma_j^z\rangle=\lim_{\theta\rightarrow\infty}\frac{\langle\psi_{\rm T}|\E^{-\theta H}\sigma_i^z\sigma_j^z \E^{-\theta H}|\psi_{\rm T}\rangle}{\langle\psi_{\rm T}|\E^{-2\theta H}|\psi_{\rm T}\rangle}.
\end{equation}
Here, the trial wave function $|\psi_{\rm T}\rangle$ is required to be 
non-orthogonal to the true groundstate wavefunction.
We take  $|\psi_{\rm T}\rangle$ as the groundstate of the non-interacting
system ($U=0$).
In the actual calculations, the projection parameter $\theta$ is chosen sufficiently large as to ensure convergence to the system's
groundstate wavefunction.
Here, we employed a value of up to $\theta=250$, which was necessary to meet this requirement.
Furthermore, we implemented a third-order symmetric Suzuki-Trotter
decomposition of the projection operator,  with an imaginary-time discretization step
${\Delta\tau=0.05}$, as well as an $SU(2)$ spin symmetric Hubbard-Stratonovich
decoupling of the local Hubbard interaction term.
Further details of the algorithm may be found in Ref.~\onlinecite{assaad_det_qmc}.

\subsection{Effective theory vs. QMC in a zigzag ribbon\label{sec_qmc_vs_eff_zigzag}}

In a zigzag ribbon one edge terminates with $A$ sublattice sites while the other edge terminates on the $B$ sublattice. As a consequence the edge states of opposite edges hybridize via the hopping Hamiltonian, while states localized at the same edge interact only via the Hubbard term. It is therefore convenient to formulate the effective theory of a zigzag ribbon in the basis of edge state operators $e_{k s \tau}$, where $k\in[\frac{2\pi}3,\frac{4\pi}3]$ labels the momentum along the edge and $s=u,l$ labels the upper and lower edge, respectively. $\tau$ is a spin label. Note that these basis states are not eigenstates of the hopping Hamiltonian. In this basis the effective Hamiltonian reads
\begin{multline}
H_{\rm eff} = \sum_{k\tau} t_k e^\dagger_{k s \tau} e^{}_{k\bar s\tau} \\+ \frac U{N_x} \sum_s \sideset{}{'} \sum_{k,k',q} \Gamma(k,k',q) :e^\dagger_{k+q s\uparrow} e^{}_{ks\uparrow}::e^\dagger_{k'-qs\downarrow}e^{}_{k's\downarrow}:.\label{eff_zigzag_hamiltonian}
\end{multline}
The primed momentum summation means that $\frac{2\pi}3\leq k+q,\,k,\,k'-q,\,k' \leq \frac{4\pi}3$. $N_x$ is the length of the ribbon, i.e., the number of unit cells along the edge. $t_k$ is the hybridization between the edge states of given momentum $k$ at different edges. It is equal to the smallest eigenvalue of the hopping Hamiltonian for $k$, which is obtained numerically.

The form of Eq. (\ref{eff_zigzag_hamiltonian}) is remarkable and deserves attention. The single-particle term is a momentum-conserving inter-edge hopping with hopping amplitude $t_k$. This is the only term that effectively couples the two edges; for all geometries considered in the present work, the inter-edge coupling via bulk states turned out to be much smaller than the direct coupling via $t_k$. The Hubbard term acts on each edge separately. Thus, the effective theory retains the bipartite character of the original honeycomb lattice. From this very form one may already expect that there is an intra-edge ferromagnetic correlation mediated directly by the Hubbard term and an antiferromagnetic inter-edge coupling in second order perturbation theory of the order of $t_k^2/U$.

\begin{figure}[!ht]
\centering
\includegraphics[width=240pt]{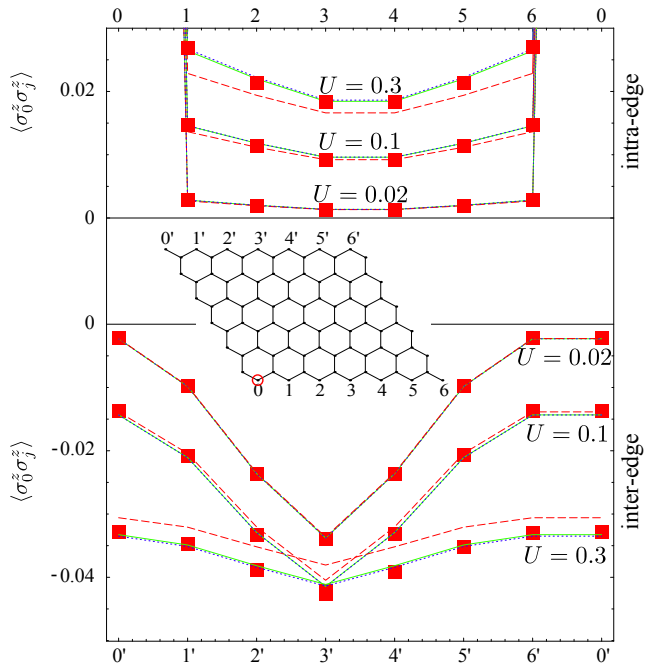}
\caption{(Color online) Spin correlation functions $\langle \sigma^z_0 \sigma^z_j\rangle$ at the edges of a zigzag ribbon ($N_x=7$, $N_y=6$) for three values of $U=0.02, 0.1,0.3$. Site 0 in the first index of the correlation function refers to site 0 at the lower edge (as indicated by the circle). The squares represent the exact results quantum Monte-Carlo simulations (error bars are smaller than the symbol size). The curves show the results of the effective theory. The lines between the data points are guides to the eye. The dashed red curve is the bare correlation function, consisting of edge, bulk and mixed terms. The solid green and dotted blue lines (mostly on top of each other) show the susceptibility correction without and with RPA, respectively. The edge sites are labeled as indicated in the inset.}
\label{fig_corr_zz_7_6_lowU}
\end{figure}

We now discuss the transition from local correlations between adjacent edges (small $U$) to strong ferromagnetic correlations along the edges, and demonstrate the agreement between QMC results and the effective theory, which is solved by exact diagonalization. Figure \ref{fig_corr_zz_7_6_lowU} shows the intra- and inter-edge correlation functions of a zigzag ribbon with length $N_x=7$ and width $N_y=6$ in the low-$U$ regime. We always use periodic boundary conditions along the edge. For $U=0$, the intra-edge correlation function is exactly a Kronecker delta function, while there is a somewhat smeared-out antiferromagnetic correlation between adjacent sites at different edges (site 0 at the lower edge and site 3 at the upper edge in the present example). These antiferromagnetic correlations decay along the ribbon edge. As $U$ becomes larger, the intra-edge ferromagnetic coupling becomes stronger and longer-ranged ferromagnetic correlations start to build up within each edge. For larger $U$ each edge hosts a super-spin $S=N_e/2$, where $N_e$ is the number of edge states with 
different momenta. The inter-edge antiferromagnetic couplings, which have a characteristic strength of $t_k^2/U$ become weaker as $U$ is increased, but they are still sufficiently strong to produce an antiferromagnetic alignment of the intra-edge super-spins in the ground state.

Note that the results of the effective model calculations with the most relevant background corrections (bulk susceptibility with and without RPA) agree very well with the exact QMC results (square symbols) for $U\lesssim 0.3$. For $U\leq0.1$ the background corrections are rather small but for larger $U$ they become more and more significant since the bulk states become locally polarized via the edge state correlations (for more details see the derivation of the background corrections in Appendix \ref{appendix_background_correction}). However, it should be emphasized that these background corrections are very short-ranged. They do not affect the long-range physics but only enhance the correlations stemming from the edge states. Solely the edge states, which are well described by the effective low-energy theory $H_{\rm eff}$, determine the long-range behavior of the correlation function.

\begin{figure}[!ht]
\centering
\includegraphics[width=240pt]{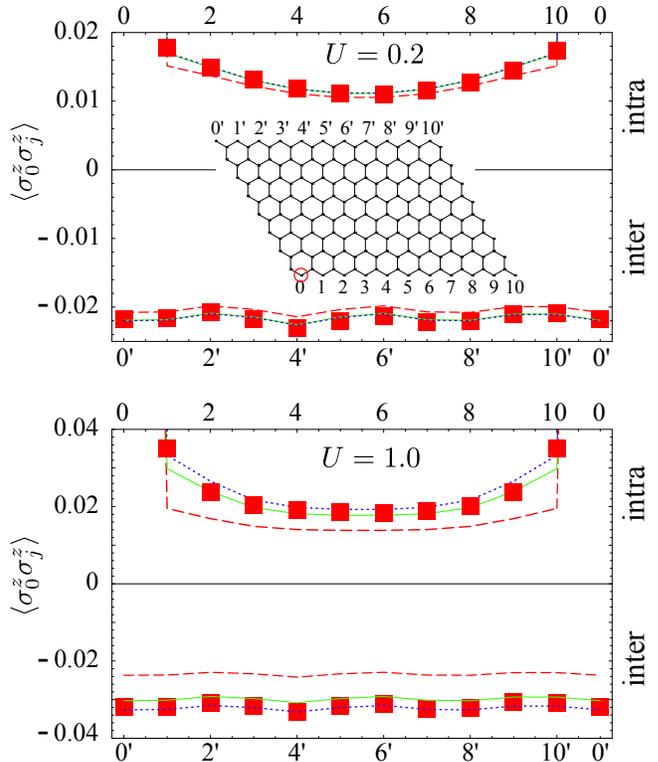}
\caption{(Color online) Spin correlation functions $\langle \sigma^z_0 \sigma^z_j\rangle$ at the edges of a zigzag ribbon ($N_x=11$, $N_y=8$) for larger $U=0.2, \, 1.0$. The meaning of the symbols and lines is the same as in Fig.~\ref{fig_corr_zz_7_6_lowU}.}
\label{fig_corr_zz_11_8}
\end{figure}

For large $U$ of the order of the nearest-neighbor hopping amplitude the background corrections are well visible (see Fig. \ref{fig_corr_zz_11_8}). In the regime where $U$ is larger than the inter-edge hopping in the effective theory, the edge state contribution to the correlation function is saturated and thus becomes independent of $U$. In this regime, only the background correction increases, but the edge state correlations are saturated. Also, for larger $U$ there is a visible difference between the non-interacting background correction $C^{(1)}$ and the RPA background correction $C^{(\mathrm{RPA})}$. Naturally, if $U$ comes too close to the RPA critical $U_c\simeq 2.23$~\cite{sorella_mean_field} the background correction is overestimated. This is why it is difficult to reproduce the exact QMC correlation function for that large $U$ as this would imply the need for high order corrections in perturbation theory. However, for understanding the underlying long range physics, which is contained in the edge state subsystem, it is not even necessary to reproduce the exact lattice correlation function with all corrections sitting on top of the actually relevant edge state physics. The study of the edge contribution plus all corrections rather obfuscates the simple physical picture that will be discussed in the next section.

\subsection{Effective theory vs. QMC in chiral ribbons}

\begin{figure}[!ht]
\centering
\includegraphics[width=\linewidth]{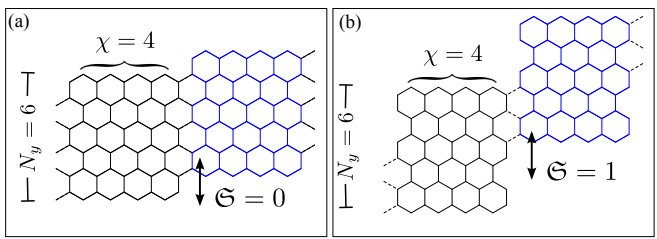}
\caption{(Color online) Definition of the shifted chirality geometry via $\chi,\,\mathfrak S,$ and $N_y$. Dashed lines indicate connections between sites of neighboring unit cells. Part (a) shows a standard chiral nanoribbon. Part (b) shows a nanoribbon with an additional shift $\mathfrak S=1$ of the unit cells.}
\label{fig_chiral_geometry}
\end{figure}

The derivation of an effective low-energy theory is not restricted to zigzag ribbons. Also for chiral ribbons we may obtain an effective edge state theory along the general lines described in Sec.~\ref{sec_derivation}. We characterize the ribbon geometry by the number of unit cells along the ribbon $N_x$, the number of zigzag lines across the ribbon $N_y$, the length of one unit cell $\chi$, and an additional shift $\mathfrak S$ of the unit cells. Figure \ref{fig_chiral_geometry} clarifies those definitions. Note that the shift $\mathfrak S$ changes the chirality of the ribbon in a non-standard way (standard would be $\mathfrak S=0$). We introduce $\mathfrak S$ in order to be able to tune the localization of an edge state along the edge (see below).

The eigenstates of $H_0$ may be obtained in $k$-space
\begin{equation}
\psi_{k\alpha}(m,n) = \frac1{\sqrt{N_x}}\, \E^{\I k m} \phi_{k\alpha}(n),
\end{equation}
where $m$ runs over different unit cells, $n$ runs through the sites within one unit cell, and $\phi_{k\alpha}(n)$ is the transverse wave function for the $\alpha$th eigenstate with momentum $k$.

\begin{figure}[!ht]
\centering
\includegraphics[width=240pt]{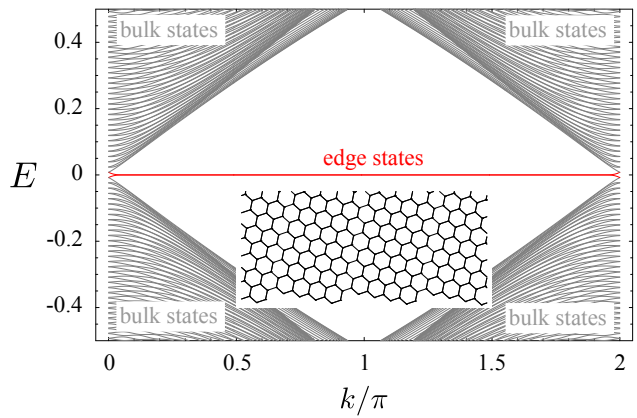}
\caption{(Color online) Edge states and bulk states of a chiral ribbon with $\chi=4$ and $N_y=200$. The inset shows the geometry along one edge.}
\label{fig_chiral_band_structure}
\end{figure}

In Fig. \ref{fig_chiral_band_structure} we show the band structure of a standard ($\mathfrak S=0$) ribbon with $\chi=4$ and $N_y=200$. Obviously there is a clear partitioning between edge states and bulk states in this chiral ribbon. In ribbons with different parameters one may find similar partitions, although the edge states may not exist for all momenta in the Brillouin zone.

\begin{figure}[!ht]
\includegraphics[width=\linewidth]{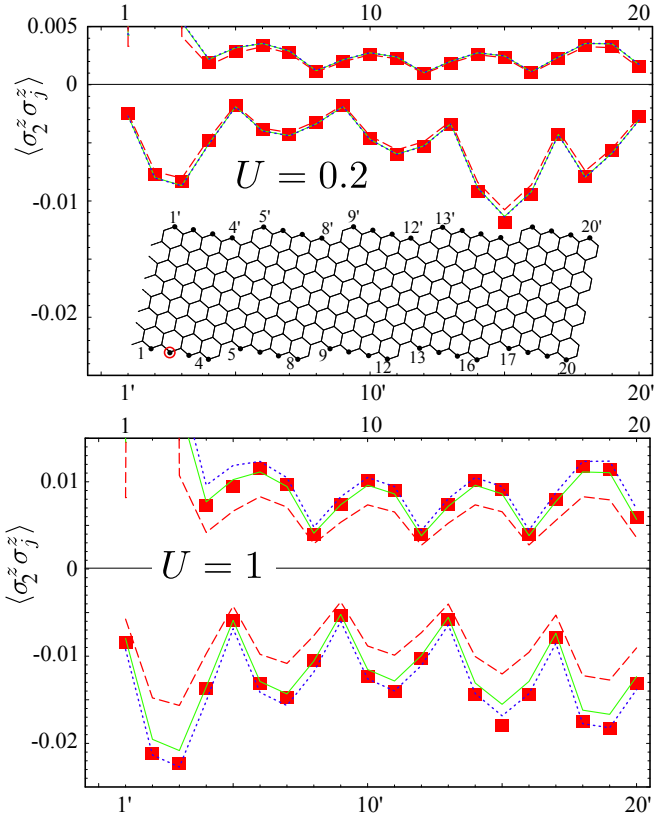}
\caption{(Color online) Spin correlation function $\langle \sigma_2^z \sigma_j^z\rangle$ at the edges of a chiral ribbon for two interaction strengths $U=0.2,\,1$. The number of unit cells is $N_x=5$ and the number of zigzag lines within each unit cell is $N_y=8$. The positive correlations are intra-edge and the negative correlations are inter-edge. The edge sites labeling is indicated in the inset. The meaning of the symbols and lines is the same as in Fig. \ref{fig_corr_zz_7_6_lowU}.}
\label{fig_corr_chiral_540}
\end{figure}

\begin{figure}[!ht]
\includegraphics[width=\linewidth]{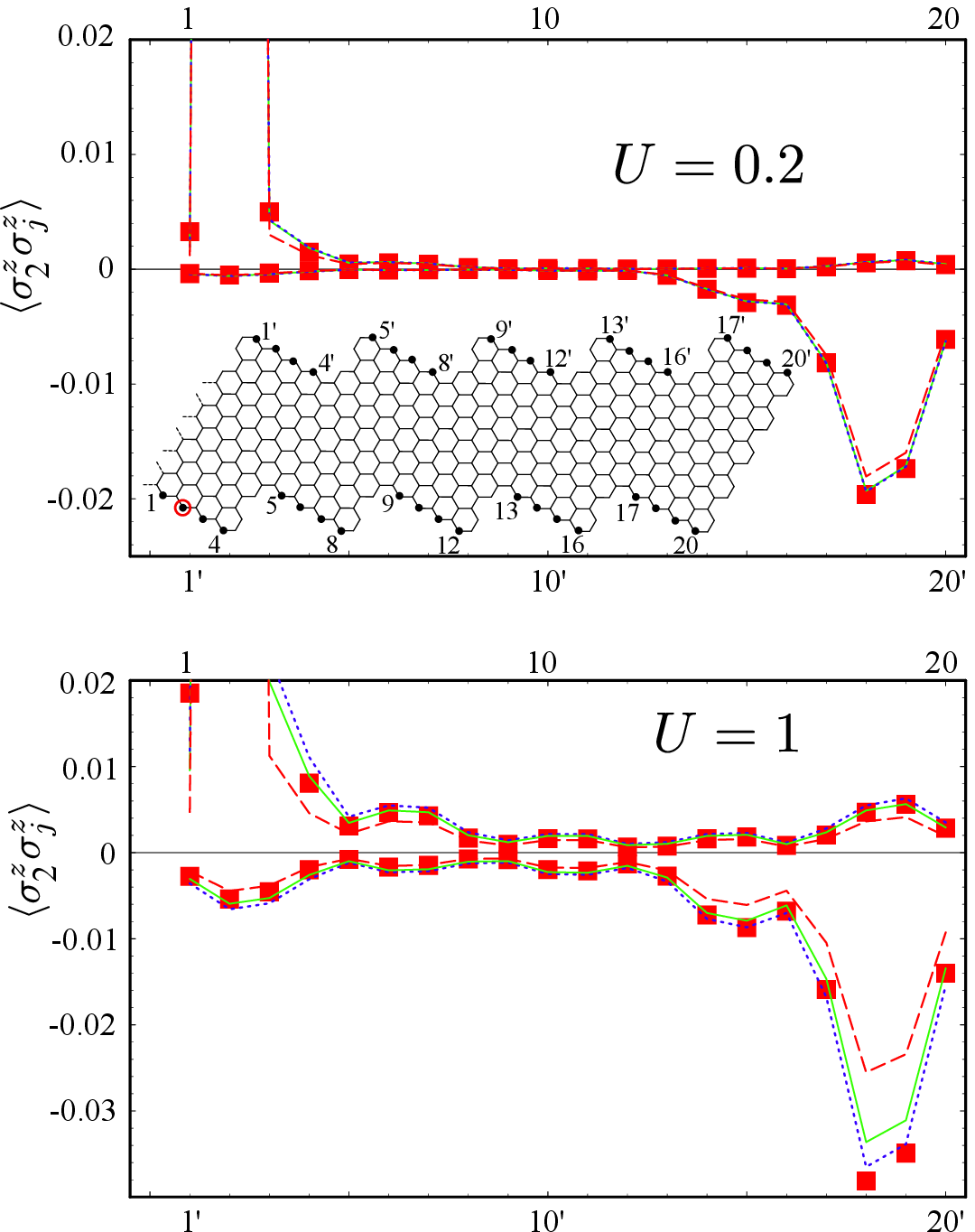}
\caption{(Color online) Spin correlation function $\langle \sigma_2^2 \sigma_j^2\rangle$ at the edges of a chiral ribbon with shift $\mathfrak{S}=1$. The number of unit cells is $N_x=5$ and the number of zigzag lines within each unit cell is $N_y=10$. The positive correlations are intra-edge and the negative correlations are inter-edge. The edge sites labeling is indicated in the inset. The meaning of the symbols and lines is the same as in Fig. \ref{fig_corr_zz_7_6_lowU}.}
\label{fig_corr_chiral_551}
\end{figure}

As for the zigzag ribbons we compare the spin correlation function $\langle \sigma^z_i \sigma^z_j\rangle$ for $i,j$ right at the edge. We do so for two chiral geometries: in one case (geometry 1) we use $N_y=8$ and $\mathfrak S=0$ (Fig. \ref{fig_corr_chiral_540}) and in the other case (geometry 2) we use $N_y=10$ and $\mathfrak S=1$ (Fig. \ref{fig_corr_chiral_551}). $N_x=5$ and $\chi=4$ in both cases. As far as the agreement of QMC and the effective theory is concerned, the result is very similar to what we have found for zigzag ribbons. For small $U=0.2$ the agreement is perfect and the background correction is very small. For larger $U=1$ we find a considerable background correction that enhances the underlying edge state correlations.

More important than a perfect reproduction of the QMC results by the effective theory is the qualitative agreement with respect to the suppression of the ferromagnetic order along the ribbon. In case of geometry 1 the ferromagnetic correlation is already strong for $U=0.2$ and for $U=1$ the edge state ferromagnetism is essentially saturated and the correlations are extended along the edges. For geometry 2, however, the ferromagnetism along the ribbon is strongly suppressed even for $U=1$. We attribute this suppression to the weakened ferromagnetic interactions due to $\mathfrak S>0$. It is most important to notice that the effective model agrees with the exact QMC solution regarding even this subtle point. 

\section{An intuitive picture\label{sec_intuitive_picture}}

The bipartite nature of the honeycomb lattice translates to a bipartiteness in the effective edge state theory. The spectrum of $H_0$ is particle-hole symmetric at half-filling, i.e., all eigenstates come in pairs of positive and negative energies $\pm\epsilon_\alpha$. Let their wave functions be $\phi_{\alpha\pm}(i)\in\mathbb R$. The two linear combinations $\phi_{\alpha+}(i) + \phi_{\alpha-}(i)$ and $\phi_{\alpha+}(i) - \phi_{\alpha-}(i)$ live on different sublattices. This principle obviously holds also for the edge states. Thus, we may choose the edge state basis $e_{\alpha s \tau}$ where $s=A,B$ labels the sublattice, on which the wave function lives. One example for this sublattice-resolved basis was used in Sec. \ref{sec_qmc_vs_eff_zigzag}, where $\alpha$ was a momentum. Another example will be studied in the next section, where $\alpha$ labels the center of a Wannier state. In this sublattice-resolved basis, $H_0$ is of the form
\begin{equation}
H_0 = \sum_{s\alpha\alpha'\tau} t_{\alpha\alpha'} e^\dagger_{\alpha s\tau}e_{\alpha' \bar s\tau},\label{zigzag_hopping}
\end{equation}
i.e., it couples only states on different sublattices. For clean zigzag ribbons the edge states on different sublattices also live on different edges. In contrast, the Hubbard Hamiltonian
\begin{equation}
H_U = U \sum_{1234,s} \Gamma_{1234,s} e^\dagger_{1 s\uparrow}e_{2 s\uparrow} e^\dagger_{3 s\downarrow} e_{4 s\downarrow} \label{zigzag_hubbard}
\end{equation}
couples only states on the same sublattice, as is obvious from the definition of the effective vertex $\Gamma$ in Eq. (\ref{gamma_def}).

Based on this bipartitioning of the edge states into groups on different sublattices one arrives at the following picture: $H_U$ essentially gives rise to a ferromagnetic coupling of order $U$ within each sublattice. On the other hand, $H_0$ gives rise to an antiferromagnetic coupling of order $|t_{\alpha\alpha'}|^2/U$ between different sublattices. This picture becomes especially clear if instead of plane waves along the edge, one uses a localized Wannier basis. This will be discussed in detail in Sec. \ref{sec_heisenberg}. For perfect zigzag ribbons it turns out that the $k$-space basis is superior to the localized Wannier basis. But for chiral ribbons and, as we expect, also for disordered ribbons, where the edge states are naturally more localized along the edge, the Wannier basis turns out to open up promising possibilities for a further simplification of the fermionic edge state model to a Heisenberg model.

For the zigzag edge states, however, we may still demonstrate the qualitative validity of the intuitive picture on a heuristic level. For this we investigate the excitation spectrum of the edge state subsystem by exact diagonalization of the Hamiltonian of the fermionic effective theory. The excitation energies of the bulk states are much higher so that they will not interfere in this low-energy consideration. In these spectra we will identify the traces of the two different couplings (intra- and inter-edge).

\begin{figure}[!ht]
\centering
\includegraphics[width=\linewidth]{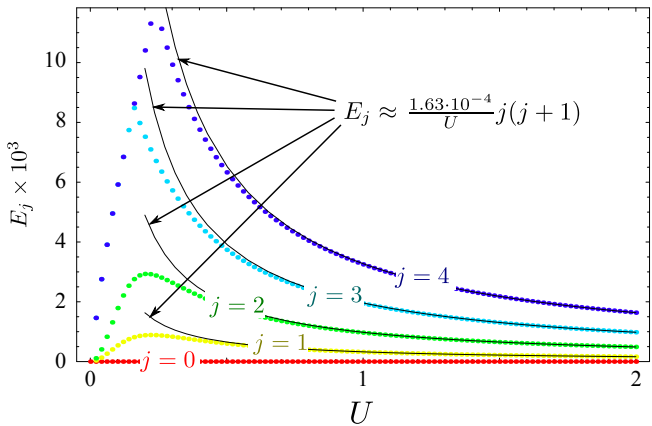}
\caption{(Color online) Lowest excitation energies of a zigzag ribbon (${N_x = 11}$, ${N_y=20}$) as a function of $U$. For orientation the ground state energy ($E=0$) is included in this plot. The dots are results from the exact diagonalization of the effective edge state Hamiltonian. The lines show the super-spin approximation for large $U$ (see text).}
\label{fig_ex_spectrum_withU}
\end{figure}

Figure \ref{fig_ex_spectrum_withU} shows the lowest four excitation energies of a zigzag ribbon with length $N_x=11$. Such a ribbon has four edge states at each edge with momenta $k=8\pi/11,10\pi/11,12\pi/11,14\pi/11$. For large $U$ the four edge states at each edge form $S=2$ super spins $\ve S_A$ and $\ve S_B$, since the intra-edge coupling mediated by $H_U$ [Eq. (\ref{zigzag_hubbard})] is dominant in this regime. The effective inter-edge hopping $H_0$ [Eq. (\ref{zigzag_hopping})] couples the super spins antiferromagnetically, which may be expressed as $J \ve S_A\cdot \ve S_B$. The total spin $\ve J = \ve S_A+\ve S_B$ is zero in the ground state and the excitation energies are $\frac J2 j(j+1)$, with $j=1,\dots,4$. In order to show that the inter-edge coupling $J$ is inversely proportional to $U$, we fit $ \frac AU j(j+1)$ to the large $U$ part of the exact diagonalization results and get very good agreement for large $U$ -- not only for single curves with fixed $j$, but also for different excited states. For 
small $U$ the ED results deviate from the super-spin approximation, which signals the breaking up of the super spins.

At this point it is easy to understand the saturating behavior of the edge state theory for large $U$. For small $U$ the antiferromagnetic coupling of edge sites at opposite edges is relevant and thus suppresses the long range ferromagnetic order along the ribbon. As $U$ is increased, the intra-edge ferromagnetic correlations become stronger. In the limit of large $U$ each edge (of finite length) is ferromagnetically ordered and constitutes a rigid super-spin. The much weaker antiferromagnetic inter-edge coupling then forces the super spins into an antiparallel alignment, but is not strong enough to destroy the super spins and therewith the ferromagnetic ground state correlations that are extended along the (finite-sized) ribbon edge.

\section{Effective Heisenberg model for chiral ribbons\label{sec_heisenberg}}

Based on the analysis given above it is tempting to develop an intuitive picture on the basis of effective spins at the (zigzag or chiral) edge, being ferromagnetically coupled along the same edge and antiferromagnetically coupled between different ribbon edges. We now demonstrate that, in order to actually obtain a satisfactory description in terms of effective spins, one needs to achieve a certain degree of spatial separation of the effective edge spins in real space. For zigzag edges the edge states are not sufficiently separated for a quantitative agreement.

\subsection{Wannier edge states}

The effective theory for edge states at perfect zigzag ribbons is best formulated in momentum space. It is of course possible to transform any momentum space basis to a real space basis, i.e., a maximally localized Wannier basis (henceforth we will call such a localized basis {\it Wannier edge states}). However for zigzag edges the Wannier edge states are rather delocalized along the edge so that neighboring wave functions have a considerable overlap.

In chiral ribbons, where the edge consists of a series of zigzag segments separated by steps (see Fig. \ref{fig_chiral_geometry}), it is possible to obtain improved Wannier edge states. Each zigzag segment hosts one localized edge state and each step acts as a barrier between neighboring zigzag segments. In order to be able to further tune the Wannier edge state separation, we will make use of the shifted chirality geometry, in which the barrier height $\mathfrak S$ may be changed by shifting the zigzag segments relative to each other. Zigzag segments of length $\chi=4$ are especially convenient as a starting point, since in the folded Brillouin zone (see Fig. \ref{fig_chiral_band_structure}) there is an edge state for each momentum $k$ along the edge (remember that the edge state momenta are usually restricted to some interval, e.g. $[2\pi/3,4\pi/3]$ for zigzag edges). 

Given the transverse wave functions $\phi_{ks}(n)$ of the edge states, where $n$ labels the site within one unit cell and $s$ labels the edge, the Wannier edge states are
\begin{equation}
\psi_{xs}(m,n) = \frac{1}{N_x} \sum_k \E^{-\I k(x-m)} \phi_{ks}(n).\label{wannier_wfkt}
\end{equation}
Here, $x$ is an integer labeling the unit cell to which the Wannier edge state is localized. $\phi_{ks}(n)$ is obtained numerically from the hopping Hamiltonian $H_0$ in k-space, and so is the hybridization amplitude $t_k$ between $\phi_{ks}$ and $\phi_{k\bar s}$. The arbitrary phase of $\phi_{ks}(n)$ is fixed by requiring $\sum_{m,n} |\psi_{xs}(m,n)|^4$ to be maximal. Thus, each pair $x,s$ corresponds to one spin-degenerate localized state. In what follows we will see that the largest part of the Hubbard interaction will force one single electron to such a state, thus generating a localized spin.

A convenient measure for the degree of localization is the ratio
\begin{equation}
R_{\rm loc} = \frac{\sum_{m,n} |\psi_{x,s}(m,n)|^4}{\sum_{m,n} |\psi_{x,s}(m,n)|^2|\psi_{x+1,s}(m,n)|^2}
\end{equation}
of the self-overlap and the overlap of neighboring Wannier edge states. The numerator of $R_{\rm loc}$ corresponds to an effective Hubbard $U^*$, i.e., the energy penalty of a double occupation of one Wannier edge state. The denominator will later be interpreted as the nearest-neighbor part of the ferromagnetic intra-edge coupling.

\begin{figure}[!ht]
\centering
\includegraphics[width=220pt]{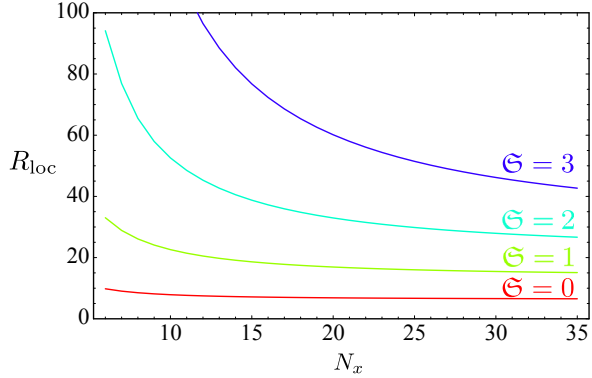}
\caption{(Color online) $R_{\rm loc}$ as a function of the ribbon geometry $N_y=2N_x$ and $\mathfrak S$, as indicated.}
\label{fig_rloc}
\end{figure}

For a zigzag ribbon, $R_{\rm loc}$ is typically between 4 and 5 (depending on the length and width of the ribbon). For chiral ribbons with variable $\mathfrak S$ the behavior of $R_{\rm loc}$ with system size is illustrated in Fig. \ref{fig_rloc}. It grows dramatically as $\mathfrak S$ is increased. Since the Heisenberg model becomes better for larger $R_{\rm loc}$ it is clear that the Heisenberg approximation is better for larger $\mathfrak S$, i.e., for larger barriers along the edge.

The effective edge state Hamiltonian can be expressed in the basis of Wannier edge states. The hopping part reads
\begin{equation}
H_{\rm hop} = \sum_{\tau,x,x'} t_{xx'} e^\dagger_{xs\tau} e_{x'\bar s\tau} + \mathrm{H.c.}\,,\label{eff_hop_chiral}
\end{equation}
with
\begin{equation}
 t_{xx'} = \frac1{N_x} \sum_{k}t_k \E^{\I k (x-x')}.
\end{equation}
$H_{\rm hop}$ couples Wannier edge states at different edges. In contrast, the effective interaction is only non-zero if all states participating in one vertex are at the same edge
\begin{equation}
H_{U} = U \sum_{1234,s} \Gamma_{x_1x_2x_3x_4,s} e^\dagger_{x_1 s \uparrow} e_{x_2 s\uparrow}e^\dagger_{x_3 s \downarrow} e_{x_4 s\downarrow}.
\end{equation}

\subsection{Heisenberg model}

As the Wannier edge states becomes more and more separated, the edge states increasingly behave like a series of Heisenberg spins. We now derive the effective couplings between those edge spins. For this we consider pairs of localized Wannier basis states and perform an up to second order perturbation theory expansion with $1/U\Gamma_{xxxx}$ as the small parameter.

We start with two states $e_x$ with $x=1,2$ at the same edge $s$, therefore sitting on the same sublattice. The effective hopping between states on the same sublattice vanishes and we are left with three terms from $H_U$ (we drop the index $s$)
\begin{align}
H_U^* &= U \sum_{x=1,2} \Gamma_{xxxx} e^\dagger_{x\uparrow} e_{x\uparrow} e^\dagger_{x\downarrow} e_{x\downarrow}\,, \\
H_J^{(1)} &= U \Gamma_{11 22} \sum_\tau (e^\dagger_{1\tau}e_{1\tau} e^\dagger_{2\bar\tau} e_{2\bar\tau} - e^\dagger_{1\tau}e_{1\bar\tau} e^\dagger_{2\bar\tau} e_{2\tau} )\,, \\
H' &= U\Gamma_{1122} e^\dagger_{1\uparrow}e_{2\uparrow} e^\dagger_{1\downarrow}e_{2\downarrow} + U\sum_\tau \Gamma_{1222}e^\dagger_{1\tau}e_{2\tau}e^\dagger_{2\bar\tau}e_{2\bar\tau} \nonumber \\
&\bs + U\sum_\tau \Gamma_{2111} e^\dagger_{2\tau}e_{1\tau}e^\dagger_{1\bar\tau}e_{1\bar\tau}+ \mathrm{H.c.}
\end{align}
The first term $H^*_U$ is a Hubbard interaction for the localized edge states. It forces the states 1 and 2 to be occupied with one electron, respectively. Since $H^*_U$ is the dominant term in the Hamiltonian it is a good approximation to restrict the Hilbert space to those manybody states having one electron per Wannier edge state. Within this restricted Hilbert space $H_J^{(1)}$ acts as a ferromagnetic coupling between the spins of the electrons in the two Wannier edge states. $H'$ has no non-zero matrix-elements in this reduced space. In principle it is straightforward to calculate the second order perturbation theory correction due to $H'$, i.e. $-H' (H_U^*)^{-1} H'$. However, for the geometries we are considering here this correction is extremely small and we were not able to find any noticeable effect of these terms on the correlation function. In the following we will drop this correction. Thus the intra-edge ferromagnetic coupling can be written as
\begin{equation}
H_\text{FM} = -\sum_{x<x',s} J_{xx',s}\boldsymbol\sigma_{xs} \cdot \boldsymbol \sigma_{x's}, \bs J_{xx',s} = \frac{U\Gamma_{xxx'x',s}}2,
\end{equation}
where $\boldsymbol \sigma_{xs}$ is the vector of Pauli matrices corresponding to the spin of the electron in Wannier edge state $x$ at edge $s$.

\begin{figure}[!ht]
\centering
\includegraphics[width=230pt]{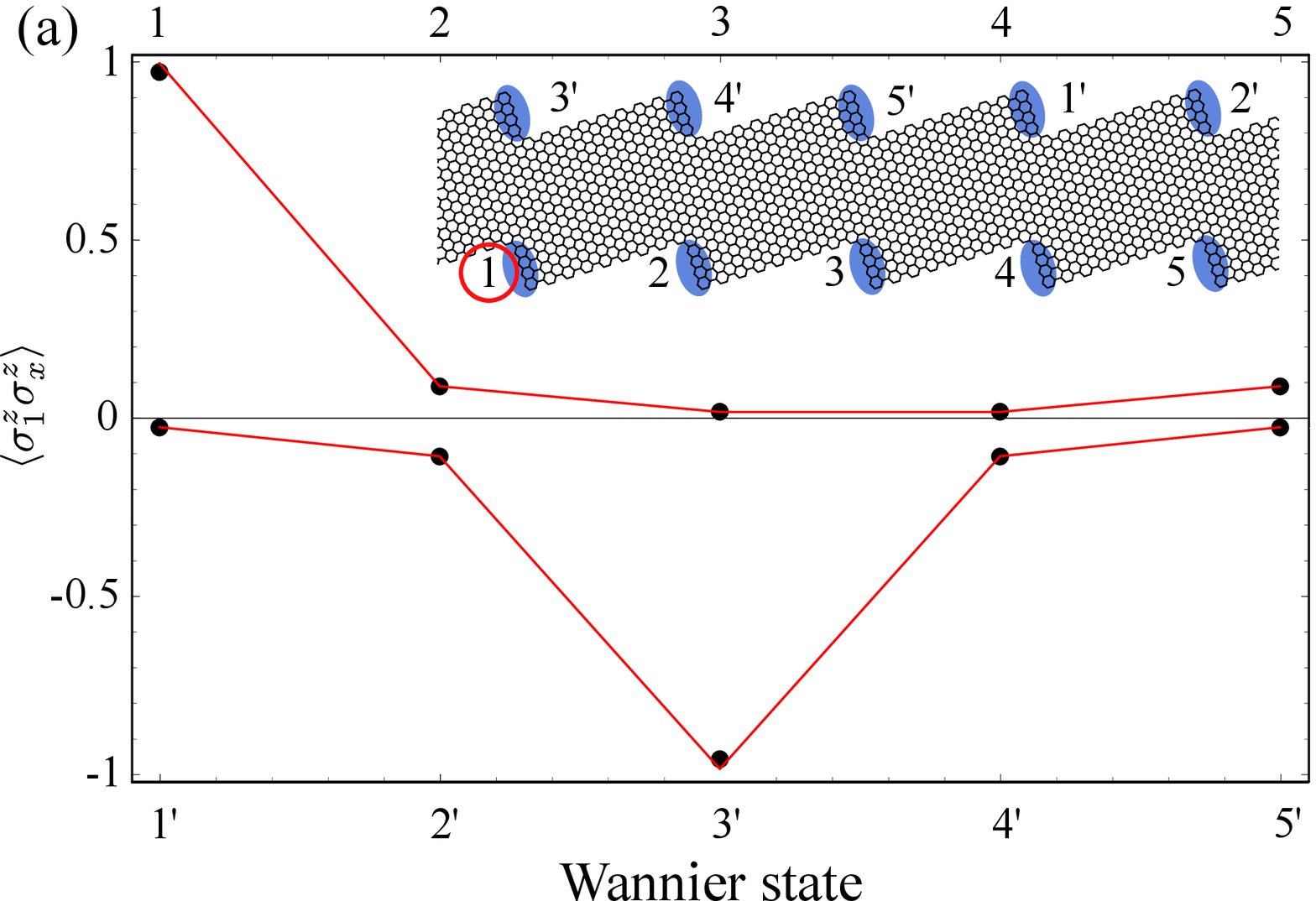}
\includegraphics[width=230pt]{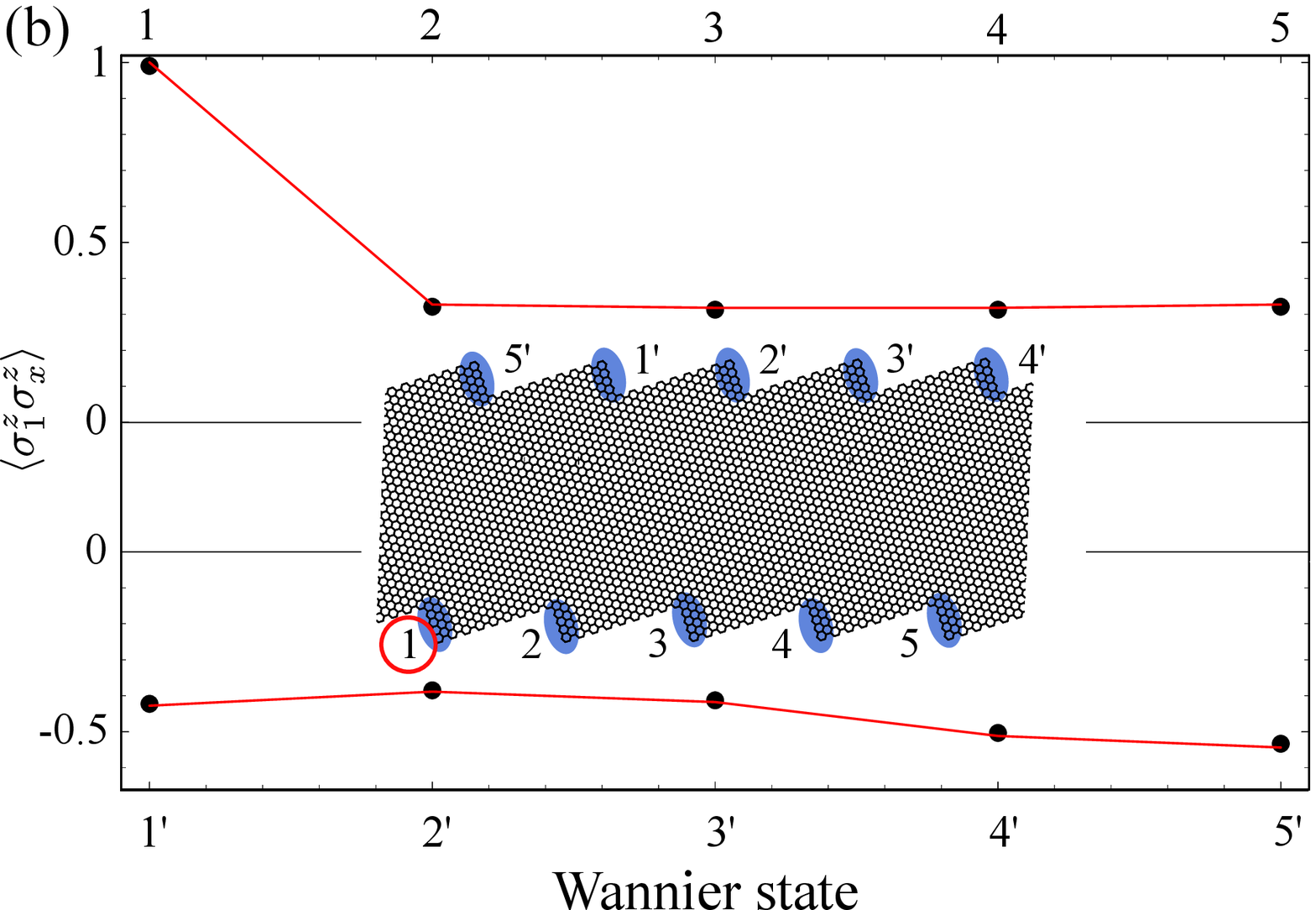}
\includegraphics[width=230pt]{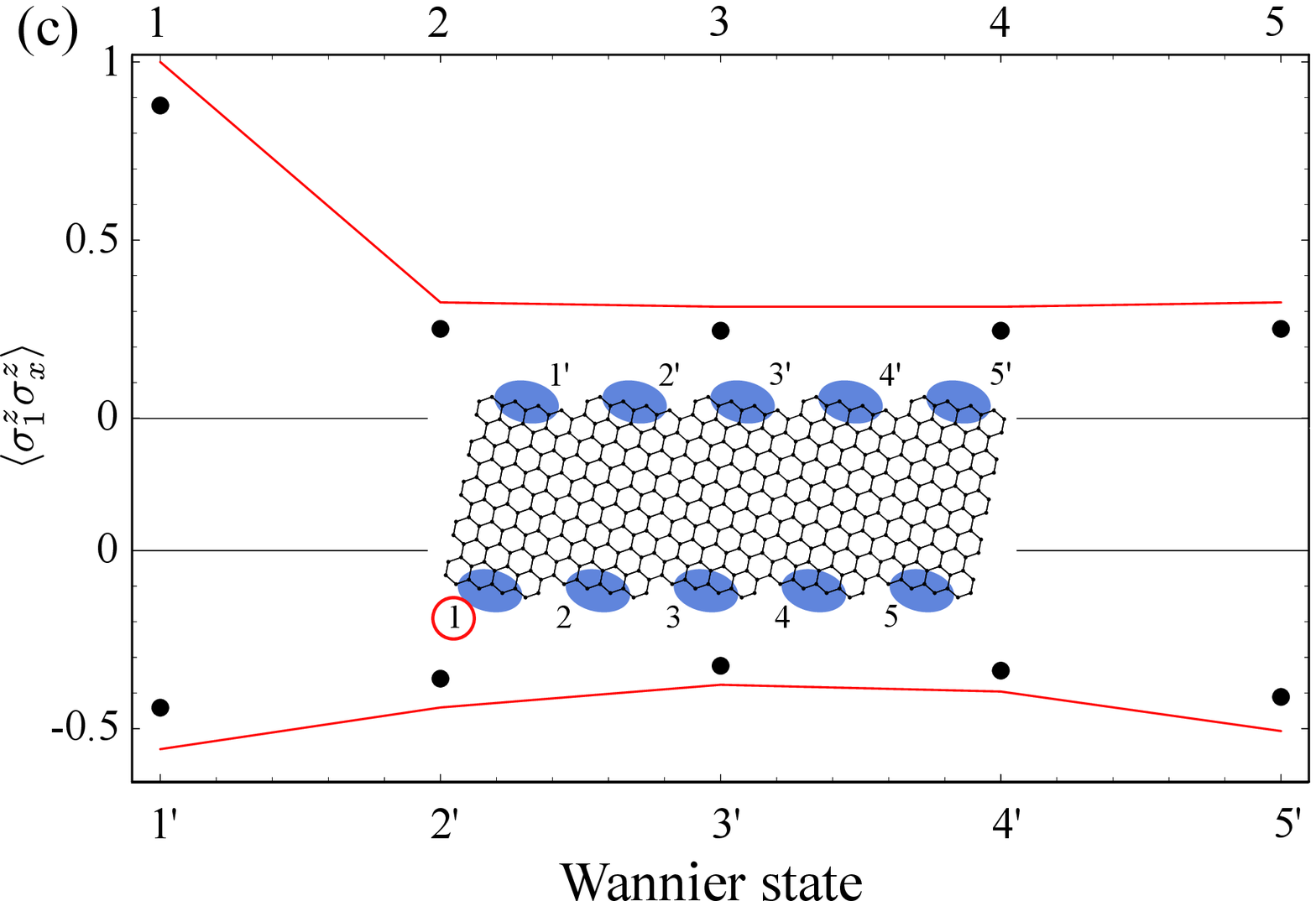}
\caption{(Color online) Effective correlation function $\langle \sigma_1^z \sigma_x^z\rangle$ of the Wannier edge states of chiral ribbons. The parameters are: $N_x=5$, $N_y=56$, and $\mathfrak S=8$ in part (a); $N_x=5$, $N_y=120$, and $\mathfrak S=8$ in part (b); $N_x=5$, $N_y=10$, and $\mathfrak S=0$ in part (c). The insets show the lattice geometry together with the positions of the Wannier states. $U=2$ in all graphs. The reference site is marked by a red circle. The positive correlations correspond to Wannier states at the same edge as the reference site. The dots are calculated from the fermionic edge state theory. The line connects points calculated from the Heisenberg model, which is an approximation of the fermionic theory. The line segments connecting the Wannier state numbers are only guides to the eye.}
\label{fig_heisenberg_vs_fermionic}
\end{figure}

Next we consider a pair of Wannier edge states $e_x$ with $x=1,2$ at different edges, therefore sitting on different sublattices. In this case there are only two terms in the Hamiltonian. The dominant term is again $H^*_U$ so that we again may restrict ourselves to the case of one electron per edge state. The second part is the effective hopping $H_{\rm hop}$ [Eq. (\ref{eff_hop_chiral})] which has no matrix elements in the restricted subspace and must therefore be taken into account in second order perturbation theory via $-H_{\rm hop} (H_U^*)^{-1} H_{\rm hop}$. In this case one obtains the standard antiferromagnetic interaction between the spins of the electrons in edge states 1 and 2, i.e.,
\begin{equation}
H_\text{AFM} = \sum_{x,x'} J_{xx'}^{s\bar s}   \boldsymbol\sigma_{xs} \cdot \boldsymbol \sigma_{x'\bar s}
\end{equation}
with
\begin{equation}
J_{xx'}^{s\bar s} = \frac{t^2_{xx'}( \Gamma_{xxxx,s} +\Gamma_{x'x'x'x',\bar s})}{2U\Gamma_{xxxx,s}\Gamma_{x'x'x'x',\bar s}} .
\end{equation}
Note that this expression also holds for edges without translation symmetry (e.g. disordered edges). For the chiral ribbons with translation symmetry we have $U\Gamma_{xxxx,s} = U^*$ for all $x,s$ and thus the expression for the antiferromagnetic coupling reduces to the well-known $J_{xx'}^{s\bar s} = t^2_{xx'}/ U^*$.

\subsection{Fermionic theory vs. Heisenberg theory}

In order to scrutinize the Heisenberg approximation we compare the correlation functions of the spins in the singly-occupied Wannier edge states, calculated within the fermionic edge state theory and in its Heisenberg approximation. In both cases the ground state is calculated by exactly diagonalizing the effective Hamiltonians. The fermionic correlation function is
\begin{equation}
\langle \sigma^z_{xs} \sigma^z_{x's'}\rangle^{\rm ferm.} = \sum_{\tau\tau'} \sigma^z_{\tau\tau} \sigma^z_{\tau'\tau'} \langle e^\dagger_{x s\tau} e_{x s \tau} e^\dagger_{x's'\tau'}e_{x's'\tau'}\rangle.
\end{equation}

The results of this comparison for three example geometries and $U=2$ is shown in Fig. \ref{fig_heisenberg_vs_fermionic}. In part (a) and part (b) the Wannier edge states are well separated due to large chiral shifts $\mathfrak S=8$, which results in a nearly perfect agreement between the two edge state theories. In part (c) a shift $\mathfrak S=0$ geometry is considered. In this case, there are apparent differences between the correlation functions calculated from the fermionic theory and from the Heisenberg approximation, which can be traced to the fact that the neighboring Wannier edge states have a relatively large overlap ($R_{\rm loc} \approx 8.4$ in this case). Still the qualitative behavior of the magnetic correlations along the edges is reproduced.

An important point to note is that the effective Heisenberg theory is able to reproduce the transition between the local inter-edge antiferromagnetic correlation [part (a) of Fig. \ref{fig_heisenberg_vs_fermionic}] and the extended intra-edge ferromagnetic correlation [part (b) of Fig. \ref{fig_heisenberg_vs_fermionic}]. The exact solution of the fermionic edge state theory via exact diagonalization is limited to $O(10)$ Wannier edge states. The effective Heisenberg spin model, however, is solvable for several thousand spins by means of highly efficient world-line QMC simulations, even in the presence of extended exchange interactions~\cite{sselongrange}. The crucial observation here is that  for  the effective Heisenberg model no QMC sign problem emerges, because the bipartiteness of the honeycomb lattice underlying the original Hubbard model description translates into a bipartitioning of the effective Heisenberg spin model, along with ferromagnetic (antiferromagnetic) Heisenberg exchange interactions among sites that belong to equal (different) sublattices. This remarkable feature reminds of a similar commensurability effect in the RKKY-interactions among magnetic adatoms or lattice-defect-induced 
local moments mediated by the graphene bulk electrons on the honeycomb lattice~\cite{rkky1,rkky2}.

\section{Summary and discussion}

We have shown that the magnetic features of graphene edges that are due to electron-electron interactions can be described solely on the basis of effective theories for the edge states. We have given simple and general rules for the construction of these effective theories, applicable to arbitrary edge geometries. The observables defined on the original honeycomb lattice, the most important of which is the spin-spin correlation function, can be reconstructed from the edge state correlation function predicted by the effective theory. This reconstruction involves a background correction from the bulk states, but only in a non-interacting or mean-field approximation, so that no elaborate methods are needed for its evaluation. In fact, one obtains already a reasonably good estimate of the background correction by simply multiplying the edge state correlation function from the effective theory by a factor of $1.5 U$. This shows that the background correction has essentially the effect of a trivial local 
amplification of the spin-correlations arising from the edge states.

Based on this observation we  argued that in order to understand the correlations along graphene edges it is advantageous to study the effective theory directly. The main reason for this is that the complicated structure of the lattice-resolved correlation function tends to obscure the underlying physics. This can be seen easily by comparing Figs. \ref{fig_corr_chiral_540} and \ref{fig_heisenberg_vs_fermionic}, describing essentially the same physics, but in the first plot the original lattice is reconstructed, while in the second plot it is not.

If the edge states can be written in a real space basis with well separated wave functions, a further approximation of the fermionic theory, namely a Heisenberg theory of edge states, is feasible and able to reproduce the effective correlation functions. We have given simple formulas by which the Heisenberg coupling constants can be evaluated directly from the wave functions of the hopping Hamiltonian. Furthermore, we have shown that the fermionic edge state theory (which agrees well with the exact solution) agrees even quantitatively with the much simpler Heisenberg theory for the edge states, as long as the edge states are well separated in their maximally localized Wannier basis. Even for less well separated edge states we still find qualitative agreement.

The effective theories described here enable the theoretical study of graphene systems of realistic sizes. Especially with the Heisenberg model it is possible to study thousands of spins, which corresponds to ribbon lengths of micrometers.

\acknowledgments

We acknowledge insightful discussions with C.~Honerkamp, C.~Koop, R.~Mazzarello, and M.~Morgenstern. Financial support by the DFG under Grant WE 3649/2-1 is gratefully acknowledged, as well as the allocation of CPU time within JARA-HPC and from JSC J\"ulich.

\appendix

\section{Background correction\label{appendix_background_correction}}

The exact zero temperature spin-spin correlation function between two sites $i,j$ on the honeycomb lattice reads
\begin{multline}
\langle\sigma^z_i\sigma^z_j\rangle = C(i,j) = \sum_{\tau\tau'}\tau\tau' \sum_{1234} \psi^*_1(i)\psi_2(i) \psi^*_3(j)\psi_4(j) \\\times\langle \Psi_0| d_{1\tau}^\dagger d_{2\tau}d^\dagger_{3\tau'}d_{4\tau'}|\Psi_0\rangle,
\end{multline}
where $d_{\mu\tau}$ annihilates an electron with spin $\tau$ in a state $\mu$ with wave function $\psi_\mu(i)$. We assume that $d_{\mu\tau}$ is an eigenstate of $H_0$ [Eq. (\ref{orig_hopping})] with eigenvalue $\epsilon_\mu$. $|\Psi_0\rangle$ is the ground state, the average with respect to which may be written as (following the standard procedure described in Ref. \onlinecite{fetter_walecka})
\begin{multline}
\langle \Psi_0| d_{1\tau}^\dagger d_{2\tau}d^\dagger_{3\tau'}d_{4\tau'}|\Psi_0\rangle \\ = \sum_{n=0}^\infty \frac{(-\I)^n}{n!} \int_{-\infty}^{\infty} \D t_1\dots \D t_n e^{-\eta(|t_1|+\dots+|t_n|)} \\ 
\times \langle\Phi_0| T[H_1(t_1)\dots H_1(t_n) d_{1\tau}^\dagger d_{2\tau}d^\dagger_{3\tau'}d_{4\tau'} ] |\Phi_0\rangle_c,\label{ex_gs_av}
\end{multline}
where $H_1(t)$ is the interaction picture operator of all terms of $H_U$ [Eq. (\ref{orig_hubbard})] but the ones involving four edge state operators. $|\Phi_0\rangle$ is the ground state of $H_0$. The terms with four edge state operators are accounted for exactly in the effective edge state theory. Note also that only connected diagrams are to be included in this series.

We may write $H_1(t)$ as
\begin{equation}
H_1(t) = U \sideset{}{'} \sum_{1234} \Gamma_{1234} d^\dagger_{1\uparrow}(t)d_{2\uparrow}(t) d_{3\downarrow}^\dagger(t) d_{4\downarrow}(t),
\end{equation}
where the primed sum means that terms with four edge state operators are excluded. Since in the only terms coupling edge and bulk states are contained in $H_1$, the bulk state operators may be contracted separately in Eq. (\ref{ex_gs_av}), while the edge state averages are left to be calculated exactly.


It turns out to be sufficient to only retain the first order in $U$ in the perturbation series (\ref{ex_gs_av}). From the corresponding average $ \langle\Phi_0| T[H_1(t) d_{1\tau}^\dagger d_{2\tau}d^\dagger_{3\tau'}d_{4\tau'} ] |\Phi_0\rangle_c$ we select only the terms in which there is one pair of edge state creation/annihilation operators in $H_1$ and one pair in the group $d_{1\tau}^\dagger d_{2\tau}d^\dagger_{3\tau'}d_{4\tau'}$. We drop all other terms since they are smaller and the quality of the background correction reached by this lowest order correction is already much better than the error bars involved in the parameters (the Hubbard $U$ used in the literature fluctuates by factors up to 4) entering the initial model. Besides it is not our aim to actually reconstruct the exact result as good as possible. We rather argue that it is in principle possible to increase the agreement with the exact correlation function in a perturbative way, but for understanding the underlying physics it is not 
recommended to work in a lattice formulation at all.

Within the approximation described above the first order correction to the correlation function reads
\begin{widetext}
\begin{multline}
C^{(1)}(i,j) \simeq U \sum_{\substack{1\dots 8\\\tau\tau'\tau_1}} \tau\tau' \left[ \psi^*_1(i)\psi_2(i) \psi^*_3(j)\psi_4(j) + \psi^*_1(j)\psi_2(j) \psi^*_3(i)\psi_4(i)\right] \sum_{i_1} \psi^*_5(i_1)\psi_6(i_1)\psi^*_7(i_1)\psi_8(i_1)\\
 (-\I) \int \D t e^{-\eta |t|} \langle  e^\dagger_{5\tau_1}e_{6\tau_1} e_{1\tau}^\dagger e_{2\tau} \rangle \langle\Phi_0| T[b^\dagger_{7\bar\tau_1}(t) b_{8\bar\tau_1} (t)  b^\dagger_{3\tau'} b_{4\tau'}]|\Phi_0\rangle_c.
\end{multline}
\end{widetext}
Remember that in our notation the operator structure determines whether the numeric indices $1\dots8$ run over bulk or edge states. In this expression we calculate the edge state correlation function $\langle  e^\dagger_{5\tau_1}e_{6\tau_1} e_{1\tau}^\dagger e_{2\tau} \rangle$ from the effective edge state theory. The remaining four point function involving the bulk state operators may be expressed in terms of the {\it bulk state spin susceptibility}
\begin{equation}
\chi^b_{ij} = \sideset{}{'}\sum_{\substack{8\text{ unocc.}\\7\text{ occ.}}}\frac{2{\rm Re}[\psi^*_8(j)\psi_7(j) \psi^*_7(i_1)\psi_8(i_1) ]}{\epsilon_8-\epsilon_7} ,
\end{equation}
where the state summation is restricted to the bulk states only. A straightforward calculation then gives
\begin{equation}
C^{(1)}(i,j) = U \sum_{i_1} \left[ \langle \sigma^z_i \sigma^z_{i_1}\rangle^e \chi_{i_1 j}^b + \chi_{i i_1}^b \langle \sigma^z_{i_1} \sigma^z_{j}\rangle^e\right].\label{background_correction}
\end{equation}

This expression for the background correction gives rise to an appealing interpretation. For this we assume that $i$ and $j$ are edge sites with a distance of, say, at least 5 lattice constants. One may understand the measurement of the spin-spin correlation function between two sites $i$ and $j$ as fixing the spin at site $i$ and then measuring the mean spin at site $j$. Fixing the spin at $i$ leads to a long-range spin polarization coming from the edge states, expressed within the effective edge state model by $\langle \sigma^z_i \sigma^z_{i_1}\rangle^e$. This polarization may be measured directly at $j$ by setting $i_1=j$. But there is also the effect that the bulk states, which are themselves not capable of developing long-range correlations, become spin-polarized near a non-vanishing edge state polarization at $i_1$. This is because the bulk states feel an effective Zeeman field $\frac U2 \langle \sigma^z_i \sigma^z_{i_1}\rangle^e$ at site $i_1$ due to the edge state spin correlations. This additional 
spin polarization of the bulk states at site $j$ due to the edge state polarization at site $i_1$ is described by the bulk spin susceptibility $\chi_{i_1j}^b$. This is the first term in Eq. (\ref{background_correction}). The second term is the symmetric process under the exchange of $i$ and $j$.

For the geometries studied in the present paper, $\chi_{ij}^b$ depends separately on $i$ and $j$. However, it is always true that the diagonal contribution $\chi^b_{ii}\simeq 0.4\pm0.1$ is by far the largest. Thus, the main effect of taking the correction $C^{(1)}(i,j)$ into account amounts to a multiplication of the edge state correlation function by $(1+0.4 U)$.

Following the argumentation above, higher order (in $U$) corrections may be taken partially into account by replacing $\chi_{ij}^b$ by its RPA series. As shown in the main part of the paper, doing so increases the agreement with the exact correlation function for $U$ below the mean-field critical interaction for the antiferromagnetic instability. For $U$ close to $U_{\rm crit.}$, however, the overestimation of the antiferromagnetic spin response by RPA naturally leads to a considerable overestimation of the total spin-spin correlation. Nevertheless, the qualitative long-range structure of the spin correlations can be determined solely within the effective theory for the edge states.

%
%

\bibliography{refs}

\end{document}